\documentclass[12pt,a4paper]{article}
\usepackage[utf8]{inputenc}
\usepackage{amsmath}
\usepackage{amsfonts}
\usepackage{amssymb}
\usepackage{graphicx}
\usepackage[left=2cm,right=2cm,top=2cm,bottom=2cm]{geometry}
\usepackage[round, numbers]{natbib}
\usepackage{longtable}
\usepackage{hyperref}
\usepackage{authblk}
\usepackage{tabularx}
\usepackage{ltablex} 

\title{Long-term predictive models for mosquito borne diseases: a narrative review}

\author[1]{Marcio Maciel Bastos}
\author[1]{Luiz Max Carvalho}
\author[1]{Eduardo Correa Araujo}
\author[1]{Flávio Codeço Coelho}
\affil[1]{Applied Mathematics School, Getulio Vargas Foundation, Rio de Janeiro, Brazil}

\date{}

\begin{document}

\maketitle

\begin{abstract}
In face of climate change and increasing urbanization, the predictive mosquito-borne diseases (MBD) transmission models require constant updates.
Thus, is urgent to comprehend the driving forces of this non stationary behavior, observed through spatial and incidence expansion.

This review explores multidisciplinary approaches to explore the advancements and challenges in long-term predictive modeling for MBDs, integrating meteorological, epidemiological, and socio-demographic factors through mathematical, statistical and computational methodologies.

We explore the role of climatic variables, such as temperature and precipitation, alongside human dynamics like population density and mobility, in shaping disease transmission as reported in the peer-reviewed literature.
This review not only address the current state-of-the-art in predictive modeling but also identifies critical gaps and innovative approaches.
We also discuss the potential for adopting modeling techniques from related fields, such as Meteorology and Economics, to improve predictive accuracy.
Finally, we address open questions and future directions related to long-term prediction and the integration of important sources of information, expanding data sources and refining models with poorly explored socio-economic and environmental variables.

We observed that temperature is a critical driver in predictive models for MBD transmission, also being consistently used in multiple reviewed papers with considerable incidence predictive capacity.
Rainfall, however, have more subtle importance as moderate precipitation creates breeding sites for mosquitoes, but excessive rainfall can reduce \textit{larvae} populations.
We highlight the frequent use of mechanistic models, particularly those that integrate temperature-dependent biological parameters of disease transmission in incidence proxies as the Vectorial Capacity (VC) and temperature-based basic reproduction number - R0(t), for example.
These models show the importance of climate variables, but the socio-demographic factors are often not considered.
This gap is a significant opportunity for future research to incorporate socio-demographic data into long-term predictive models for more comprehensive and reliable forecasts.

With this survey, we outline the most promising paths to be followed by long-term MBD transmission research and highlighting the potential facing challenges.
Thus, we offer a valuable foundation for enhancing disease forecasting models and supporting more effective public health interventions, specially in the long term.
\end{abstract}

\section{Introduction}
\label{sec:intro}

    Mosquito-borne diseases (MBDs) represent some of the most significant public health challenges globally \citep{WHO2020Vector}.
    Dengue alone is responsible for more than 3.9 billion infected annually in over 129 countries at dengue risk of contracting, with estimated 96 million symptomatic cases and 40,000 deaths every year \citep{WHO2020Vector}.
    The rapid proliferation and geographical expansion of diseases such as dengue, chikungunya and Zika points to the need of urgent effort in order to better understand the drivers of their spread. 
    Thus, understanding the role of extrinsic variables such as climate, especially temperature and pluviometry, demographic and social dynamics in the modulation of the transmission mechanism of MBD, offers crucial insights into vectors and pathogens' spread. 
    
    In what follows, we will analyze a selection of prominent studies on long-term predictions and related ideas with the potential to be applied to models of this nature, elucidating their methodologies, key findings and broader implications. 
    Through a detailed exploration of the literature, our intention is to offer a rigorous synthesis on the state-of-the-art of mathematical and computational techniques of MBD prediction and analysis, as well as an immediate path forward for the development of models with decadal prediction (forecasting over a ten-year period).
    Decadal predictions are important to provide insights for long-term planning and policymaking, anticipating optimal resource usage and enable proactive measures to mitigate risks and damages to public health.

    Each selected article embodies a distinct facet of MBD research, together they present a comprehensive overview of the myriad analytical tools currently available.
    This review intends to map the different approaches of these methodologies, along with potential gaps and improvements.
    
\section{Methodology}
\label{sec:methods}

    In conducting a narrative review of models for long-term incidence inference of mosquito-borne diseases (MBDs), we employed a systematic search strategy utilizing two primary literature databases: PubMed and Google Scholar.
    The search was designed to identify relevant studies published from 2010 onward.
    Our search terms included``Mosquito-Borne Diseases",``Decadal Models",``Long term incidence models",``MBD forecasting",``MBD time series analysis",``Quantitative analysis of mosquito-borne diseases",``MBD predictive models",``Weather",``El Niño" and``Population dynamics".

    To ensure the relevance and quality of the selected articles, we applied specific inclusion and exclusion criteria. Articles were included if they were published from 2010 onward, written in English and included a quantitative analysis. 
    Additionally, studies had to be directly related to long-term predictions of MBD incidence.
    We excluded articles with significant methodological biases or dubious findings, identified by the presence of unsubstantiated claims or lack of methodological rigor. 
    Studies not pertinent to the topic, such as those focusing on unrelated diseases or short-term predictions, were analyzed to consider possible long-term expansions, if they were related to other long-term models studied or were simply removed from the scope.
    Due to the limited number of studies on long-term predictions of MBD incidence, we included in our analysis studies with prediction periods exceeding two years. 
    This adjustment aligns with the shorter period of the El Niño phenomenon.
    To enrich this narrative review, we included long-term prediction studies from related fields such as demography, sociology and weather forecasting.
    These fields offer valuable methodologies and insights that can improve the predictive accuracy and interdisciplinary understanding of MBD incidence. 
    
    For each study, we examined the following aspects:  objectives, methodology(ies) employed, data sources, key findings, strengths and limitations.
    
\section{A Review of Existing Models}
\label{sec:review}

We have broken down our analysis by the prediction horizon of the studies: we begin with a review of the existing long-term models and then move on to describe shorter-term prediction models which we believe could be appropriately extended in order to provide longer-term predictions.

\subsection{Long-term Models for Mosquito-Borne Disease Incidence}
\label{sec:long_term}

    In one of the earliest works on long-term MBD prediction, \citet{patz1998dengue} explore the relationship between climate change and dengue fever transmission,  applying the vectorial capacity ($VC$) equation, as defined by the equation \eqref{eq:vc}, with temperature forecasts from three climate general circulation models (GCMs) in a computer simulation framework.
    
    In equation (\ref{eq:vc}), $m$ is the female mosquito population per person (vector density), $b$ is the probability of disease transmission from an infectious mosquito to a susceptible human, $c$ is the probability of a mosquito acquiring infection from an infected human, $a$ is the number of bites on humans per mosquito per day, $p$ is the daily survival rate of mosquitoes and $n$ is the duration of the extrinsic incubation period (EIP) in days.
    The study focuses on temperature as the primary climate variable affecting dengue transmission dynamics through the parameters of the vectorial capacity equation (VC). The authors chose not to include other climate variables, such as precipitation, because while rainfall can influence mosquito breeding sites and larval development, its relationship with adult mosquito survival and dengue transmission is less direct and varies across regions. 
    This approach explains the future epidemic potential (EP) of dengue, defined as the reciprocal of the critical density threshold ($m$) when $VC =1$.
    The study uses historical climate data from 1931 to 1980, specifically the temperature as the key variable influencing the parameters of the vectorial capacity equation (\ref{eq:vc}).

    \begin{equation}\label{eq:vc}
        \begin{aligned}
            VC = \frac{m \cdot b \cdot c \cdot a^2 \cdot p^n}{-\ln{(p)}}.
        \end{aligned}
    \end{equation}
    
    Their projections were based on an anticipated 1.16°C temperature increase by 2050 and reveal a possible 31-47\% rise in $EP$ of dengue in areas currently at risk, suggesting that a smaller mosquito population could maintain or enhance virus spread.
    However, the study faces limitations such as potential overestimation in hyperendemic regions and underestimation in temperate zones.
    Additionally, it overlooks crucial factors such as precipitation, mosquito species diversity and virus strains, all of which significantly influence transmission dynamics \cite{patz1998dengue}.
    \citet{ryan2019global} examine the future transmissibility of \textit{Aedes}-borne viruses (dengue, chikungunya and Zika) under climate change scenarios, utilizing General Circulation Models (GCMs).
    They applied GCMs to an empirically tailored temperature-dependent function to predict shifts in the distribution and risk associated with \textit{Aedes aegypti} and \textit{Aedes albopictus} by 2050 and 2080. 

    The basic reproduction number ($R_0$) -- a key metric in epidemiology -- represents the average number of new infections one individual can cause in a wholly susceptible population, indicating a disease's transmission potential.
    When $R_0 < 1$, there is little to no epidemic potential and when $R_0 > 1$ the disease has potential to cause widespread epidemics, as described by \citet{diekmann1990definition}.
    The study by \citet{ryan2019global} utilizes a temperature-dependent version of $R_0(T)$, as defined by the function in \eqref{eq:R0_temperature}, based on a modified Ross-Macdonald model [\citet{gorgas1916ronald}, \citet{macdonald1957epidemiology}], to analyze how temperature influences disease spread through \textit{Aedes} mosquitoes.
    Except for $N$ and $r$, which represent human population size and human recovery rate, respectively, this model includes temperature's effect on critical mosquito and virus characteristics such as the mosquito biting rate ($a$), vector competence ($b \cdot c$), mosquito mortality rate ($\mu$), viral development rate within the mosquito ($PDR$), daily egg production by female mosquitoes ($EDF$), the survival probability of mosquitoes from egg to adult ($p_{EA}$) and the development rate of mosquito from egg to adult ($MDR$).

    \begin{equation}\label{eq:R0_temperature}
        \begin{aligned}
            R_0(T) = \left\{ \frac{a(T)^2 \cdot b(T) \cdot c(T) \cdot \exp\left[-\frac{PDR(T)}{\mu(T)}\right] \cdot EFD(T) \cdot p_{EA}(T) \cdot MDR(T)}{N \cdot r \cdot \mu(T)^3} \right\}^{1/2}
        \end{aligned}
    \end{equation}
    
    Their model anticipates a poleward migration of these vectors, driven by their thermal preferences.
    This results in varying human exposure risks under different climate change scenarios, with significant regional differences in transmission risk patterns.
    The study's strength lies in its empirically calibrated model validated against actual case data, although it recognizes limitations including the uncertainties in climate modeling, the exclusion of other factors affecting transmission besides temperature and the use of a constant population from the year 2015 predicted by the \textit{Gridded Population of the World}, version 4 (GPW4).
    
    \citet{petrova2019sensitivity} investigate the use of long-lead forecasts of El Niño events to predict dengue fever outbreaks. They develop a forecasting model that exceeds traditional El Niño predictability limits by analyzing subsurface processes and stored heat, allowing for predictions up to two years in advance.
    This model, combined with a statistical dengue model, demonstrates potential for early warning of dengue epidemics. 

    Their study employed a combinations of an El Niño Southern Oscillation (ENSO) forecasting model and a dengue simulation model.
    The ENSO forecasting model was a structural time series model, formulated by the equation \eqref{eq:ENSO_pred}, where $y_t$ represents the monthly Niño index at time $t$, $\mu_t$ is the trend component, modeled as a random walk process and $\gamma_t$ is the seasonal component.
    Moreover, $psi_{1t}$, $\psi_{2t}$, $\psi_{3t}$ are three cycle components with different frequencies, persistence and variances, $x'_t\delta$ represents explanatory regression variables such as surface and subsurface ocean temperature and wind stress and $\epsilon_t$ is the irregular term accounting for random error.
    This model employs a state-space framework and the Kalman Filter for dynamic parameter estimation. 

    \begin{equation}\label{eq:ENSO_pred}
        \begin{aligned}
            y_t = \mu_t + \gamma_t + \psi_{1t} + \psi_{2t} + \psi_{3t} + x'_t\delta + \epsilon_t.
        \end{aligned}
    \end{equation}
    
    The dengue simulation model was a negative binomial generalized linear mixed model (GLMM) expressed as the distribution \eqref{eq:dengue_negbin} and equation \eqref{eq:dengue_linpred}, where $y_t$ is the monthly dengue cases, $k$ is the overdispersion parameter representing excess variance relative to the mean, $\mu_t$ is the expected number of cases, $e_t$ represents the population offset, $\alpha$ is the intercept, $\beta_{t'(t)}$ accounts for temporally autocorrelated random effects for each month, $\gamma_j x_{jt}$ the effects of climatic variables, $\delta_{T'(t)}$ represents random year effects and $\epsilon_j z_{jt}$ includes the effects of non-climate factors.
    The parameters were estimated using Markov Chain Monte Carlo (MCMC) methods within a Bayesian framework.

    \begin{equation}\label{eq:dengue_negbin}
        \begin{aligned}
            y_t \sim \operatorname{NegBin}(\mu_t, k)
        \end{aligned}
    \end{equation}

    \begin{equation}\label{eq:dengue_linpred}
        \begin{aligned}
            \log(\mu_t) = \log(e_t) + \alpha + \beta_{t'(t)} + \sum_j \gamma_j x_{jt} + \delta_{T'(t)} + \sum_j \epsilon_j z_{jt}
        \end{aligned}
    \end{equation}
    
    The model presented in \citet{petrova2019sensitivity}, while pioneering, has limited applicability due to its focus on the relationship between ENSO and dengue incidence specifically within Ecuador.
    
    \citet{munoz2021spatiotemporal} examined the spatiotemporal dynamics of dengue in Colombia in relation to local climate and ENSO, employing linear and nonlinear causality methods.
    Initially, lagged cross-correlations were utilized to assess the strength and direction of linear relationships between the dengue incidence and the climatic as well as ENSO variables over various time lags, identifying potential delayed effects of climatic changes and ENSO phases on the transmission dynamics of dengue. 
    
    Subsequently, the study uses Wavelet analysis to examine the time-frequency characteristics of dengue and Oceanic Nino Index (ONI) series and Wavelet Coherence and Cross-Wavelet Transform to help identify periods and frequency bands where the two series are related. 
    
    Also, the author employed methods including  PCMCI (Peter and Clark Momentary Conditional Independence), ParCorr (Partial Correlation), GPDC (Gaussian Process Distance Correlation), CMI (Conditional Mutual Information) to causal discovery in both linear and non-linear causal relationships among time-lagged weather variables, ENSO indices and dengue incidence.

    Their analysis revealed periodicity in ONI and dengue cases in Colombia, particularly a 32-month cycle that aligns with ENSO frequencies.
    This cycle was most pronounced between 2009 and 2013 and is associated with the strong link between ENSO events and dengue incidence in the country.
    The maximum cross-correlations between ONI and climatic variables show that El Niño typically leads to drier periods and warmer temperatures across most of Colombia, with Andes, Pacific and Caribbean being the regions more affected by ENSO. 
    Warmer temperatures can accelerate the life cycle and increase the biting rates of the Aedes aegypti mosquito. However, precipitation is negatively correlated with the number of dengue cases in Colombia at the national scale and varies across regions. 
    Also, the authors discovered that higher rainfall tends to reduce dengue incidence and reduced precipitation during El Niño can lead to fewer mosquito breeding sites initially, but subsequent concentrated breeding in available water sources can result in a higher density of mosquitoes.

    \citet{martheswaran2022prediction} developed three specific variations of the Susceptible-Infected-Removed (SIR) model to analyze dengue case data from Singapore and Honduras.
    The first is a null model, which assumes a constant transmission rate $\beta$ throughout the year, disregarding any seasonal or climatic influences on dengue transmission dynamics.
    The second variation, a seasonal model, incorporates a seasonally varying transmission parameter $\beta$, modeled using trigonometric functions as described by the equation \eqref{eq:trig_beta}.
    In this model, $\beta_0$, $\beta_1$ and $\beta_2$ are coefficients to be estimated, $t$ represents time and $T$ is the period corresponding to one year, reflecting changes in transmission rates due to seasonal environmental patterns.
    The third variation, a climate model, explicitly uses climate variables to adjust the transmission rate parameter, aiming to capture the immediate effects of weather conditions on dengue transmission.
    The parameters were estimated with \textit{Markov Chain Monte Carlo} (MCMC) technique, enabling a generative Bayesian inference, recovering the posterior distribution of the parameters instead of a point estimation.
    
    \begin{equation}\label{eq:trig_beta}
        \begin{aligned}
            \beta(t) = \beta_0 + \beta_1 \cos\left(\frac{2\pi t}{T}\right) + \beta_2 \sin\left(\frac{2\pi t}{T}\right).
        \end{aligned}
    \end{equation}

    The study compared the seasonal and climate models against the null model to highlight the significance of incorporating periodic complexity and environmental data to enhance dengue prediction.
    Specifically, it found that the seasonal and the climate models each performed better under different circumstances, with their effectiveness varying across different outbreak years and locations.
    By using the Southern Oscillation Index (SOI) and the ONI, which are utilized to explore the impact of El Niño events on dengue transmission, it was found that the El Niño variables did not  contribute to the improvement of the model's predictive ability.    
    Despite its promising outcomes, the study acknowledges limitations related to the simplification of dengue serotype dynamics and the potential underestimation of non-environmental factors such as urbanization and human mobility.
    
    The study of \citet{colon2023projecting} aimed to forecast the incidence and burden of dengue in Southeast Asia until 2099, considering climate change, urbanization and mobility.
    The authors employed a Generalized Additive Mixed Model (GAMM) with a conditional negative binomial distribution to analyze the incidence of dengue across eight Southeast Asian countries over time, incorporating spline functions for continuous predictors which allow the model to capture non-linear effects of these variables on dengue incidence.
    The model included six control variables: air temperature, number of consecutive dry days (a proxy for breeding site availability), human population density, human mobility, air travel volume and GDP.
    To account for unmeasured factors that could influence dengue risk, the model included random effects.
    The model was fitted using monthly dengue case counts at the provincial level from January 2000 to December 2017 and its predictive ability was assessed through mean absolute error (MAE), Spearman's rank correlation and blocked cross-validation.

    This model considers that dengue epidemic will follow the same trend present in the historical data, not accounting for factors as population immunity and demographic changes, which could lead to overestimation in future projections, specially in the time range proposed by the authors.

    \citet{van2023long} employed a mechanistic compartmental model to study the impact of climate change on Zika and dengue transmission in Brazil.
    The model is based on temperature-dependent coupled differential equations set describing the flow between the population compartments.
    The human population was allocated in the Susceptible $S_h$, latently infected $L_h$, infectious $I_h$ and recovered $R_h$ compartments.
    Similarly, the mosquito population was divided into susceptible $S_m$, latently infected $L_m$ and infectious $I_m$ compartments.
    The temperature-dependent parameters for \textit{Aedes aegypti} included the biting rate $a(T)$, egg-laying rate $\epsilon(T)$, survival probability from egg to adult $\theta(T)$, adult mortality rate $\mu_m(T)$ and extrinsic incubation rate $\sigma_m(T)$.
    Additionally, the model integrated the mosquito-to-human ratio $\left(\frac{N_m}{N_h}\right)$ and assumes fixed rates for recovery $(\gamma)$ and birth/Death $(\mu_h)$ within the human population.
    
    Their results indicate a general increase in the potential for epidemics under different climate scenarios, though the impact varies by region. However, the model limits to rely solely on temperature data and not incorporate other important climatic factors such as humidity and rainfall. Additionally, the model assumes fixed rates for recovery, birth and death, with the latter two being particularly critical and unrealistic for long-term predictions.

    We summarize our findings in Tables \ref{tab:pred_objectives} and \ref{tab:pred_limitations}.
    The first table provides a comprehensive overview of the objectives, methodologies, data sources, and key findings of various studies on the prediction of mosquito-borne diseases (MBDs) presented above. The second table lists the strengths and limitations of each study, providing insights into the robustness of their methodologies and the contexts in which their predictions are most applicable.

    \begin{longtable}{|p{2cm}|p{3cm}|p{3.5cm}|p{2.5cm}|p{3.5cm}|p{3cm}|p{3cm}|}
    \caption{Summary of Predictive Models for Mosquito-Borne Diseases: Objectives, Methodologies, Data Sources and Key Findings}\label{tab:pred_objectives}\\
    \hline
    \textbf{Paper} & \textbf{Objective} & \textbf{Methodology} & \textbf{Data Sources} & \textbf{Key Findings} \\
    \hline
    \endfirsthead
    
    \multicolumn{5}{c}%
    {{\bfseries \tablename\ \thetable{} -- continued from previous page}} \\
    \hline \textbf{Paper} & \textbf{Objective} & \textbf{Methodology} & \textbf{Data Sources} & \textbf{Key Findings} \\ \hline
    \endhead
    
    \hline \multicolumn{5}{r}{{Continued on next page}} \\ \hline
    \endfoot
    
    \hline
    \endlastfoot   
    
    \citet{patz1998dengue} & Examine potential added risk posed by global climate change on dengue transmission & Simulation analysis using VC equation and GCM temperature outputs & Historical climate data (1931-1980), GCM outputs & Projected temperature increase could raise epidemic potential (EP) by 31-47\% by 2050 \\
    \hline
    \citet{ryan2019global} & Predict future transmissibility of \textit{Aedes}-borne viruses under climate change scenarios & Empirically tailored temperature-dependent function with GCM applications & Climate model outputs, empirical data & Shifts in distribution and risk associated with Aedes vectors by 2050 and 2080 \\
    \hline
    \citet{munoz2021spatiotemporal} & Explore spatiotemporal dynamics of dengue in relation to climate and ENSO in Colombia & Linear and nonlinear causality methods, Wavelet analysis, PCMCI method & Local climate data, ENSO indices & Identified 32-month cycle linking ENSO events with dengue incidence \\
    \hline
    \citet{colon2023projecting} & Forecast incidence and burden of dengue in Southeast Asia under climate change scenarios & Generalized Additive Mixed Model (GAMM) with conditional negative binomial distribution & Dengue case data (2000-2017), GCM-SSP combinations & Projections indicate increased dengue burden with climate change, urbanization and mobility \\
    \hline
    \citet{van2023long} & Study impact of climate change on Zika and dengue transmission in Brazil & Mechanistic compartmental model with temperature-dependent differential equations & Local climate and epidemiological data & Increased potential for epidemics under different climate scenarios, varying regional impacts \\
    \hline
    \citet{petrova2019sensitivity} & Predict dengue outbreaks in Ecuador using long-lead forecasts of El Ni\~{n}o events & Structural time series model for ENSO forecasting, negative binomial GLMM for dengue simulation & ENSO indicators, dengue case data & Long-lead El Ni\~{n}o forecasts as early precursors of dengue epidemics, successful coupling of ENSO and dengue models \\
    \hline
    \end{longtable}
    
    \newpage
    \begin{longtable}{|p{2cm}|p{3cm}|p{3.5cm}|p{2.5cm}|p{3.5cm}|p{3cm}|p{3cm}|}
    \caption{Summary of Predictive Models for Mosquito-Borne Diseases: Strengths and Limitations}\label{tab:pred_limitations} \\
    \hline
    \textbf{Paper} & \textbf{Strengths} & \textbf{Limitations} \\
    \hline
    \citet{patz1998dengue} & Integrates climate change scenarios, well-validated VC model, broad geographical application & Potential overestimation in hyperendemic regions, underestimation in temperate zones, excludes precipitation \\
    \hline
    \citet{ryan2019global} & Uses multiple GCMs, empirical calibration & Limited to temperature-dependent factors, static population data usage \\
    \hline
    \citet{munoz2021spatiotemporal} & Combines multiple analytical techniques, strong linkage between ENSO and dengue & Focuses on specific region (Colombia), does not account for non-climatic factors \\
    \hline
    \citet{colon2023projecting} & Comprehensive model including multiple socioeconomic and climatic factors, uses ensemble GCMs & Limited to Southeast Asia, potential model overfitting \\
    \hline
    \citet{van2023long} & Detailed mechanistic approach, temperature-dependent parameters & Excludes climatic elements other than temperature, assumes fixed rates for recovery and birth/death \\
    \hline
    \citet{petrova2019sensitivity} & Provides very early warning of potential epidemics, Bayesian framework for parameter estimation & Constrained to Ecuador, lacks detailed predictive accuracy metrics \\
    \hline
    \end{longtable}
    
    \subsection{Short-term Models that could be extended to Long-term Forecasts}
    \label{sec:short_term}
    
    Transition from short-term prediction models to longer-term ones in mosquito-borne virus transmission models is challenging. This requires not only a deep understanding of existing models but also significant adaptations to account for evolving epidemiological and environmental factors. In this section we will present short-term prediction models capable of being adapted to long-term forecasts in the context of mosquito-borne virus transmission. We will also exhibit modern approaches integrating human mobility, climate change and socio-economic trends that could be used to enhance the accuracy of long-term forecasts and help to track covariates in disease dynamics.

    Mechanistic models, such as those developed by \citet{mordecai2017detecting}, 
    proved to be a useful tool for long-term predictions, not only being directly used as in \citet{ryan2019global}, but also influencing the development of more detailed models, such as in \citet{martheswaran2022prediction}.
    These models analyzed how traits and behaviors of the vectors were influenced by temperature, based on laboratory measurements. 
    Other research has adapted these mechanistic models to explore the behavior of other state variables in response to temperature variation. 
    For example, \citet{villena2022temperature}, adapted the model described by \citet{mordecai2017detecting} and developed an equation for the number of susceptible individuals $S(T)$ as a function of temperature.
    \citet{pena2023arbovirus} compared three distinct mechanistic approaches for his studies: initially, the method proposed by \citet{mordecai2017detecting}, followed by the function developed by \citet{liu2014vectorial} and finally, the approach by \citet{caminade2017global}. 
    Through the integration of these methods, the author was able to refine his predictions effectively, employing a weighted regularization strategy that mitigated the inherent biases of each model, making the predictions more reliable, though this strategy still needs further exploration to confirm its effectiveness. 
    By applying reparametrizations as done by \citet{villena2022temperature} on established long-term methods like those of \citet{ryan2019global} and \citet{martheswaran2022prediction}, it is possible to address irregularities in the state space and potentially improve inference accuracy. Additionally, both \citet{ryan2019global} and \citet{martheswaran2022prediction} based their models on the foundational work of \citet{mordecai2017detecting}. Incorporating alternative mechanistic formulations, such as those proposed by \citet{liu2014vectorial} and \citet{caminade2017global}, could enhance long-term models through balanced regularization strategies, as demonstrated by \citet{pena2023arbovirus}.

    The compartmental model, originally conceptualized pioneering by\citet{gorgas1916ronald} and \citet{macdonald1957epidemiology}, was formally introduced in its seminal form by \citet{kermack1927contribution}, known as the SIR model.
    This innovative framework divides the population into three distinct compartments, Susceptible (S), Infected (I) and Recovered (R), to capture the dynamics of disease spread within homogeneous segments of a population.
    Through a set of differential equations, the SIR model quantifies the transitions between these states, providing insights into how diseases propagate and the conditions for outbreaks.
    Moreover, models of this sort can be used to assess the effectiveness of interventions in a straightforward manner. 

    Compartmental models, in their various formulations, offer a highly effective framework for elucidating and simulating the long-term epidemiological dynamics among individuals stratified into distinct compartments.
    This approach enables a nuanced understanding of disease spread within and between these groups over time.
    As the models developed by \citet{martheswaran2022prediction} and  \citet{van2023long} exemplifies, this model framework seamlessly fit to the long-term problem solution.
    The model developed by \citet{gimenez2022vector} incorporates dynamic elements into the vector population within the compartmental framework, enhancing the model's realism and applicability to real-world scenarios.
    This approach allows for the inclusion of projections regarding changes in the vector population due to climatic shifts, urbanization and alterations in human behavior, thereby improving the accuracy of long-term forecasts.
    Furthermore, the model can be refined by relaxing the assumption of a constant host population, a simplification that often does not hold true for periods exceeding a decade. 
    This modification acknowledges the critical impact of population changes, including growth, migration, urban development and land use on disease transmission patterns.
    It is improbable to achieve high precision on long-term compartmental methods without account for these impacts provided by population dynamics, thus the technique used by  \citet{gimenez2022vector}, highlights necessary modifications that traditional compartmental method should pursue to address more realistic disease mechanism.

    The study by \citet{soriano2020vector} models the transmission of vector-borne diseases through human mobility using Monte Carlo simulations within complex networks, applying Markovian equations to capture transmission dynamics in a metapopulation framework.
    The network is structured as a series of interconnected patches (nodes) representing different geographic areas, linked by paths of human movement (edges), reflecting a realistic pattern of population distribution and mobility.
    An opportunity for model enhancement lies in incorporating detailed topological data and dynamic elements, such as public transport models or adjusting the network to reflect changes in mobility patterns over time.
    By integrating scale-free network characteristics, the model could offer more accurate epidemic forecasts.
    Adding or removing edges and nodes in response to shifts in human behavior or infrastructure developments could dynamically adjust the model, improving its utility for long-term epidemiological predictions and intervention strategies.
    Just like the methodology used by \citet{gimenez2022vector} that accounted for human dynamics, the method presented by \citet{soriano2020vector} can be integrated with other traditional methods, such as compartmental models, to develop a more realistic approach, especially using non-stationary mobility networks.
    
    Modern adaptations of mathematical models, as highlighted by \citet{tatem2013systematic}, focus on depth, emphasizing host-vector interactions and a broad array of transmission mechanisms.
    This multidisciplinary approach emphasizes the necessity of translating theoretical models into practical, actionable insights.
    For long-term predictions, this implies a need to integrate models with real-world data on climate change, urban development and socio-economic trends.
    Such an endeavor would not only improve the predictive capabilities but also align the models with public health strategies and policy-makings.
    
    Communicating the inherent uncertainties in these models, especially when extending their scope to long-term predictions, remains a significant challenge.
    As pointed out by \citet{soriano2020vector} and \citet{de2022epidemiological}, acknowledging the range of possible scenarios and effectively conveying probabilistic information is crucial.
    This transparency is key to maintaining trust and utility in epidemiological forecasting.
    Furthermore, identifying vulnerabilities and differential outcomes, as discussed by \citet{munoz2020ae} and \citet{kaur2022artificial}, is important to enhance the accuracy of predictive models.
    Long-term predictions must account for the variable susceptibilities across different populations and regions, necessitating a detailed approach that considers the different vectors, pathogens and human communities involved.

    In conclusion, extending short-term models to long-term forecasts in MBD transmission requires significant adaptations to incorporate dynamic epidemiological and environmental factors. Some models can be directly modified using similar methodologies applied to long-term predictions. Additionally, these approaches must emphasize the importance of aligning models with real-world data, addressing uncertainties and considering local singularities. This is crucial for effectively communicating probabilistic information and developing proactive public health strategies.

\section{Key Variables in Long-term Predictive Modeling}

    A comprehensive understanding of vector-borne diseases such as Zika and dengue across various studies reveals the utilization of a wide array of variables.
    These encompass human and mosquito population dynamics, climatic influences and socio-environmental factors, illustrating the complex nature of disease transmission and the importance of interdisciplinary approaches in epidemiology.
    
    Variables related to susceptible, latently infected, infectious and recovered humans are foundational in modeling the disease dynamics within human populations, as seen e.g. \citet{martheswaran2022prediction}, \citet{van2023long} and \citet{gimenez2022vector}.
    This approach is paralleled in the vector populations through variables delineating the infection status of mosquitoes, which is crucial for understanding the vector component of disease transmission.
    
    Temperature-dependent parameters for mosquitoes, including the biting rate, egg-laying rate, survival probability from egg to adult, adult mortality rate and extrinsic incubation rate, are emphasized in \citet{patz1998dengue}, \citet{ryan2019global}, \citet{martheswaran2022prediction}, \citet{gimenez2022vector} and \citet{van2023long}.
    These variables highlight the significant impact of climate on mosquito behavior and physiology, which is critical for assessing vector-borne disease transmission in the context of global warming.
    
    Climatic and environmental variables such as temperature, precipitation, relative humidity, wind velocity and ENSO indices (ESPI, OSI and ONI) are central to the studies by \citet{petrova2019sensitivity}, \citet{munoz2021spatiotemporal}, \citet{martheswaran2022prediction}, \citet{colon2023projecting}, investigating the influence of local and global climate on disease transmission patterns.
    The recurring significance of temperature and precipitation, underlining the overarching role of environmental conditions in disease ecology, is evident in their works.
    
    Furthermore, socio-economic and demographic variables, including human population density, human mobility, air travel volume and GDP can be integrated into a single analysis as done in e.g. \citet{colon2023projecting}.
    These variables underscore the complex interplay between human demographics, movement and economic status and their influence on the spread of diseases, emphasizing the human dimensions of epidemiological research.
    
    Specific variables such as the mosquito-to-human population ratio (\citet{van2023long}), recovery rate (\citet{gimenez2022vector}, \citet{martheswaran2022prediction}) and vector competence (\citet{ryan2019global}), where employed by  applied, play vital roles in assessing the potential for disease outbreaks and the effectiveness of control measures.
    These variables, critical to the specific models discussed in the cited studies, further illustrate the diverse methodologies employed to capture the multifaceted nature of disease transmission dynamics.
    
    The variables mentioned above and used in these studies are summarized in Table \ref{tab:01}, which arbitrarily group the variables into classes including climate, environmental, socioeconomic, human, mosquito and pathogen variables, along with the corresponding studies that employ them.
    
    \begin{longtable}{|p{5cm}|p{1cm}|p{1cm}|p{1cm}|p{1cm}|p{1cm}|p{1cm}|}
    \caption{Variables and Corresponding Studies} \\
    \hline\label{tab:01}
    \textbf{Variable} & \textbf{\cite{patz1998dengue}} & \textbf{\cite{ryan2019global}} & \textbf{\cite{munoz2021spatiotemporal}} & \textbf{\cite{colon2023projecting}} & \textbf{\cite{van2023long}} & \textbf{\cite{petrova2019sensitivity}} \\
    \hline
    \endfirsthead
    \multicolumn{7}{c}%
    {{\bfseries \tablename\ \thetable{} -- continued from previous page}} \\
    \hline
    \textbf{Variable} & \textbf{\cite{patz1998dengue}} & \textbf{\cite{ryan2019global}} & \textbf{\cite{munoz2021spatiotemporal}} & \textbf{\cite{colon2023projecting}} & \textbf{\cite{van2023long}} & \textbf{\cite{petrova2019sensitivity}} \\
    \hline
    \endhead
    \hline \multicolumn{7}{r}{{Continued on next page}} \\ \hline
    \endfoot
    \hline
    \endlastfoot  
    \multicolumn{7}{|c|}{\textbf{Climate Variables}} \\
    \hline
    Temperature & X & X &  & X & X &  \\
    \hline
    Minimum Temperature ($\tau_{min}$) &  &  & X &  &  & X \\
    \hline
    Maximum Temperature ($\tau_{max}$) &  &  & X &  &  &  \\
    \hline
    Precipitation &  &  & X & X &  & X \\
    \hline
    Number of Consecutive Dry Days (NCDD) &  &  &  & X &  & \\
    \hline
    Relative Humidity (H) &  &  & X &  &  &  \\
    \hline
    Wind Velocity ($v_w$) &  &  & X &  &  &  \\
    \hline
    Oceanic Niño Index (ONI) &  &  & X &  &  & X \\
    \hline
    ENSO Precipitation Index (ESPI) &  &  & X &  &  & X \\
    \hline
    \multicolumn{7}{|c|}{\textbf{Environmental Variables}} \\
    \hline
    Urbanization &  &  &  & X &  &  \\
    \hline
    \multicolumn{7}{|c|}{\textbf{Socioeconomic Variables}} \\
    \hline
    Human Population Density ($N_h/m^2$) &  &  &  & X &  &  \\
    \hline
    Human Mobility (HM) &  &  &  & X &  &  \\
    \hline
    Air Travel Volume (ATV) &  &  &  & X &  &  \\
    \hline
    GDP &  &  &  & X &  &  \\
    \hline
    \multicolumn{7}{|c|}{\textbf{Human Variables}} \\
    \hline
    Historical Case Data &  &  &  & X &  & X \\
    \hline
    Human Population ($N_h$), Susceptible ($S_h$), Infectious ($I_h$), Removed ($R_h$) &  &  &  &  & X &  \\
    \hline
    Latently Infected Humans ($\bar{I}_h$) &  &  &  &  & X &  \\
    \hline
    Human Recovery Rate ($\gamma$) &  &  &  &  & X &  \\
    \hline
    \multicolumn{7}{|c|}{\textbf{Mosquito Variables}} \\
    \hline
    Mosquito (Female) Population ($N_m$) & X &  &  &  & X &  \\
    \hline
    House Index Infestation Rate ($\delta_I$) &  &  &  &  &  & X \\
    \hline
    Mosquitoes Susceptible ($S_m$), Infectious ($I_m$) &  &  &  &  & X &  \\
    \hline
    Latently Infected Mosquitoes ($\bar{I}_m$) &  &  &  &  & X &  \\
    \hline
    Survival Probability from Egg to Adult ($S_{ea}$) &  & X &  &  & X &  \\
    \hline
    Mosquito Egg-to-Adult Development/Birth Rate ($\delta_{m}$) &  & X &  &  & X &  \\
    \hline
    Eggs Produced per Female Mosquito ($n_e$) &  & X &  &  & X &  \\
    \hline
    Biting Rate ($\delta b$) & X & X &  &  &  &  \\
    \hline
    Vector Competence ($V_c$) &  & X &  &  &  &  \\
    \hline
    Mosquito Survival/Mortality Rate (MSR/MMR) & X & X &  &  & X & \\
    \hline
    Extrinsic Incubation Period/Rate (EIP/EIR) & X & X &  &  & X & \\
    \hline
    \multicolumn{7}{|c|}{\textbf{Pathogen Variables}} \\
    \hline
    Circulating dengue Serotypes ($\sigma$) &  &  &  &  &  & X \\
    \hline
    \end{longtable}

\begin{figure}[ht]
        \centering
        \includegraphics[width=1\linewidth]{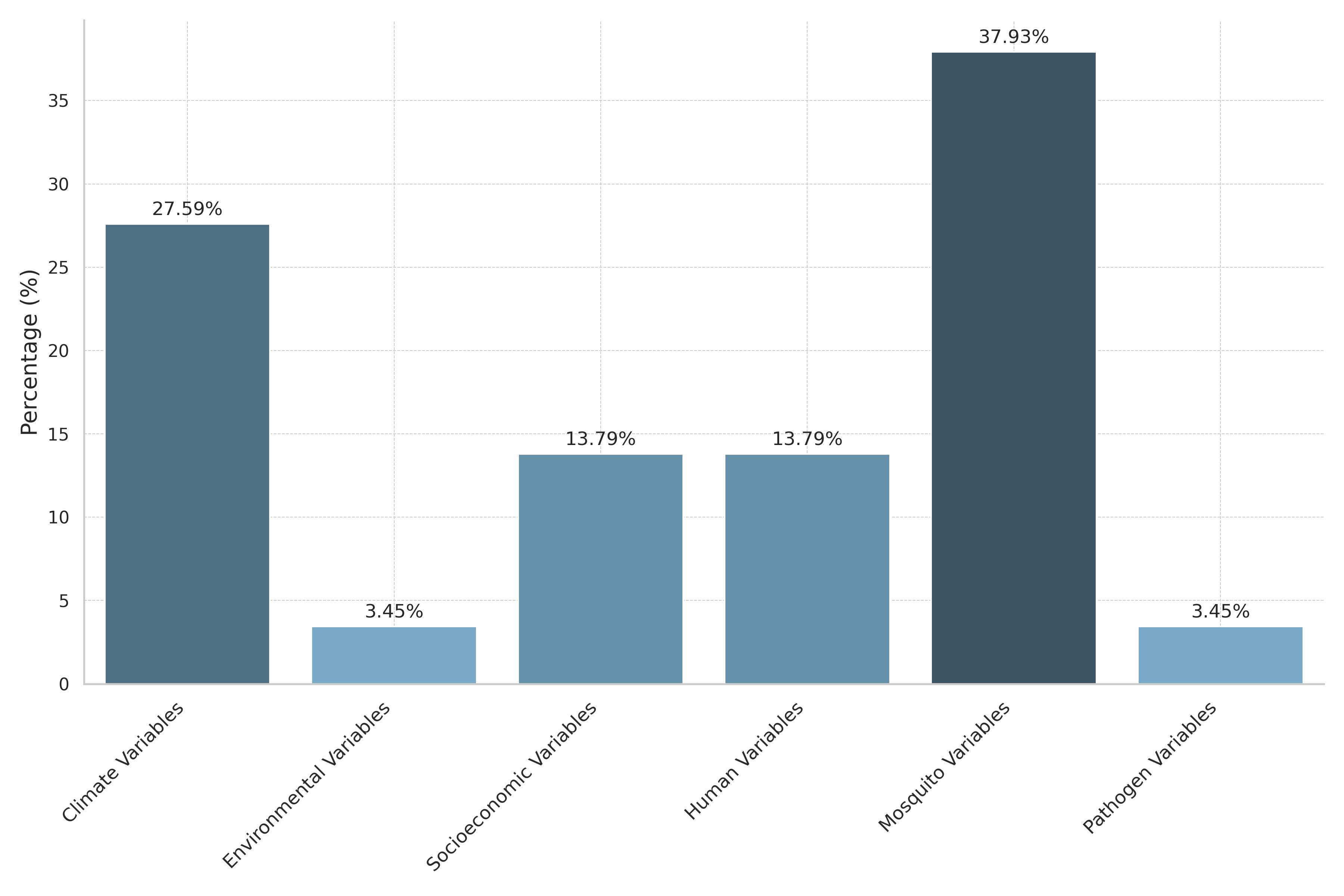}
        \caption{Frequency of Usage of Variable Categories in Predictive Models}
        \label{fig:histogram}  
    \end{figure}

    Considering the categories used to subdivide the variables listed in Table \ref{tab:01}, we can visualize and assess the popularity of these categories through the frequency histogram shown in Figure \ref{fig:histogram}.
    
\newpage
\section{Data Sources}

    Studying mosquito-borne diseases such as Zika and dengue requires researchers to gather and analyze a wide dataset, highlighting disease incidence, environmental conditions and population dynamics.
    These data, coming from different institutions and are collected using different methodologies, constitute the core of epidemiological research that aims to understand the complexities of disease transmission.

    The \citet{soriano2020vector} research provided a model integrating demographic data from Cali's municipality with mobility data from urban commuting surveys to construct a comprehensive mobility network. This approach, grounded in the collection of mobility patterns and demographic distribution, is augmented by vector distribution data sourced from the Secretaria de Salud Pública Municipal de Cali. Together, these data sources enable exploration of how human movement intersects with mosquito distribution to leverage the spread of dengue.

    \citet{martheswaran2022prediction}'s investigation of dengue fever outbreaks employs dengue incidence data from the Ministry of Health Infectious Disease Bulletin of Singapore and the Pan American Health Organization (PAHO), alongside climate data from entities such as the National Oceanic and Atmospheric Administration ( NOAA).
    This blend of health and environmental data, underpinned by robust collection methodologies including health surveillance reports and climate monitoring, highlights the pivotal role of integrating disease incidence with climatic factors to predict outbreaks.
    Similarly, \citet{gimenez2022vector} underscores the value of synthesizing data from literature reviews and public health records to parameterize epidemiological models, illustrating a methodology reliant on secondary data analysis to understand vector population dynamics and their impact on disease spread.

    In their study on the impact of El Niño on dengue epidemics in Ecuador, \citet{petrova2019sensitivity} utilize health surveillance data from Ecuador's Ministry of Health and climate data from the National Institute of Meteorology and Hydrology of Ecuador (INAMHI).
    This strategic employment of national health and meteorological data illuminates the direct correlation between specific climatic events and the incidence of dengue, showcasing the critical importance of local institutional data in disease modeling.

    Expanding the research canvas, \citet{ryan2019global} draw upon global climate data from WorldClim and population data from the Gridded Population of the World (GPW) to forecast the future risk of \textit{Aedes}-borne virus transmission under climate change scenarios.
    Their methodology leverages global datasets to model disease risk on a worldwide scale, demonstrating the application of international data sources in predicting global health patterns.

    Lastly, the work of \citet{van2023long} on the effects of temperature changes on Zika and dengue in Brazil utilizes temperature data from the Inter-Sectoral Impact Model Intercomparison Project (ISIMIP) and the Geophysical Fluid Dynamics Laboratory Earth System Model 4 (GFDL-ESM4).
    This approach, focusing on climate modeling to assess disease risk, highlights the significance of advanced climate simulations in understanding the nuanced impacts of global warming on vector-borne diseases.

    These studies demonstrate the integral role of varied data sources—from local health departments to global climate databases—in epidemiological research is evident and are summarized in the table \ref{tab:data_source}.
    Disease incidence data along with  demographic, climatic and vector distribution information, provides a comprehensive picture necessary for modeling disease transmission.
    The methodologies employed in gathering and analyzing this data, ranging from direct health surveillance to sophisticated climate projections, underscore the complex interplay between environmental, biological and human factors in shaping the epidemiological landscape of mosquito-borne diseases.

    \begin{longtable}{|p{8cm}|p{0.8cm}|p{0.8cm}|p{0.8cm}|p{0.8cm}|p{0.8cm}|p{0.8cm}|}
    \caption{Data Sources, Institutional Sources and Corresponding Studies} \\
    \hline\label{tab:data_source}
    \textbf{Data Source and Institutional Source} &     \textbf{\cite{patz1998dengue}} & \textbf{\cite{ryan2019global}} &   \textbf{\cite{munoz2021spatiotemporal}} & \textbf{\cite{colon2023projecting}}     & \textbf{\cite{van2023long}} & \textbf{\cite{petrova2019sensitivity}} \\
    \hline
    Geographic and Demographic Data (Municipality of Cali) & X &  &  &  &  &  \\
    \hline
    Mobility Data (Urban Commuting Surveys) & X &  &  &  &  &  \\
    \hline
    Vector Distribution Data (Secretaria de Salud Pública Municipal de Cali) & X    &  &  &  &  &  \\
    \hline
    Theoretical Modeling and Analysis (Literature Review, Public Health Records)    &  & X &  &  &  &  \\
    \hline
    Ministry of Health Surveillance System (Ecuador) &  &  & X &  &  &  \\
    \hline
    National Institute of Meteorology and Hydrology of Ecuador (INAMHI) &  &  & X   &  &  &  \\
    \hline
    National Service for the Control of Vector-Borne Diseases (Ecuador) &  &  & X   &  &  &  \\
    \hline
    National Reference Center (NRC) for dengue and other Arboviruses (Ecuador) &    &  & X &  &  &  \\
    \hline
    National Oceanic and Atmospheric Administration (NOAA) &  &  & X & X &  & X \\
    \hline
    Dengue Incidence and Climatic Data Integration (INS Colombia, IDEAM, NOAA) &    &  &  &  & X &  \\
    \hline
    Climate Data (WorldClim, GCMs, GPW v4) &  &  &  & X &  &  \\
    \hline
    Population Data (Gridded Population of the World, GPW v4) &  &  &  & X &  &     \\
    \hline
    Historical and Future Temperature Data (ISIMIP, GFDL-ESM4) &  &  &  &  &  & X   \\
    \hline
    Shared Socioeconomic Pathways (SSP) Scenarios (ISIMIP) &  &  &  &  &  & X \\
    \hline
    \end{longtable}

\section{Interdisciplinary Insights}
    Predicting mosquito-borne diseases requires a deep understanding of both biological and environmental factors. Advances in climatological, environmental, and socio-political forecasting offer valuable tools for enhancing the realism and accuracy of predictive models by incorporating more explanatory features. This section explores how these advances in forecasting techniques and the use of sophisticated statistical methods can be applied to improve long-term predictions and extend short-term predictions of MBD incidence.

    \citet{choi2022seasonal} developed a forecasting methodology that integrates climatological data from multiple climate models of the Coupled Model Intercomparison Project phases 5 and 6 (CMIP5 and CMIP6).
    The dataset comprised retrospective predictions (hindcasts) that have been initialized annually from the winter of 1960/1961 to 2009/2010 (50 initializations per model) each with three to ten ensemble realizations, totaling 142 ensemble instances.
    This enabled to assess seasonal-to-decadal climate phenomena, specifically on the El Niño–Southern Oscillation (ENSO) and the Pacific Decadal Oscillation (PDO).
    The forecasting approach focused on multi-model ensemble (MME) method, where outputs from different models were averaged to mitigate individual model biases and errors. Also, bias correction is performed against the observed climatology from the same period, adjusted for each lead time to counter model drift.
    The prediction skill was evaluated using the anomaly correlation coefficient (ACC) and the mean-squared skill score (MSSS).
    The results indicated that the MME approach successfully predicted ENSO with significant skill up to a year in advance and the PDO over lead times of five to nine years.
    These predictions were particularly robust at shorter leading times for ENSO and more dependent on both model initialization and the accurate estimation of near-term radiative forcing effects for longer-term PDO forecasts.
    The robustness and statistical significance of these prediction skills were assessed using a non-parametric bootstrap resampling method.
    This method involves generating thousands of synthetic hindcasts by resampling the original ensemble, allowing for the empirical estimation of confidence intervals and significance testing without assuming prediction error normally distributed.

    For long-term models like those of \citet{patz1998dengue} and \citet{van2023long}, which utilize temperature-dependent parameters, the bias correction techniques and ensemble averaging from \citet{choi2022seasonal} can be applied to improve the accuracy of temperature forecasts used in their vectorial capacity and compartmental models. Also, accurate temperature predictions can be directed applied in some of the short-term models presented, such as the temperature-dependent $R_0(T)$ function used by \citet{ryan2019global} and the temperature-specific biting rates in \citet{mordecai2017detecting}, providing a more reliable source of long-term predicted temperature data. 
    Moreover, long-term MBD models can be equipped with the use of non-parametric bootstrap resampling for assessing prediction skill to empirically estimate confidence intervals and significance levels, enabling more reliable predictions measured reports, providing valuable information for public health interventions.

    \citet{foster2020bayesian} introduces a method for forecasting sea surface temperature (SST) anomalies at high latitudes.
    Employing a Bayesian framework, the study tackles the challenge of parameter estimation within the context of Linear Inverse Models (LIMs).
    These models are mathematically structured around stochastic differential equations, with parameters \(\boldsymbol{B}\) (drift matrix) and \(\boldsymbol{Q}\) (noise covariance matrix) that are not predefined and must be estimated from the data.
    Bayesian methods, incorporating prior distributions and Monte Carlo integration techniques, facilitate the estimation of these parameters by regularizing the high-dimensional parameter spaces and integrating prior knowledge about the parameters’ statistical properties.
    The research utilizes an extensive 1800-year preindustrial control run from the Community Earth System Model version 1 (CESM1) Large Ensemble project.
    This dataset is particularly valuable as it provides a robust basis for validating the Bayesian method's effectiveness, free from the confounding effects of modern-day atmospheric forcing.
    The data is processed through Empirical Orthogonal Function (EOF) decomposition to reduce its dimensionality, focusing on ten primary patterns that explain a substantial portion of the variance in SST anomalies.
    Validation of the forecasting method was carried out by comparing Bayesian and traditional maximum likelihood estimates in terms of their ability to recover true parameter values and predict future SST anomalies accurately.
    The Bayesian approach demonstrated good performance in terms capturing the probabilistic nature of the forecast distribution, underscoring its potential to enhance regional climate predictability significantly.

    The insights from \citet{choi2022seasonal} and \citet{foster2020bayesian} are instrumental in enhancing long-range forecasts of mosquito-borne disease (MBD) incidence in a given area by leveraging the predictable components of climate variability over decadal scales.
    Forecasting the effects of sea surface temperature (SST), ENSO and PDO provides a strategic tool to anticipate climatic conditions that could significantly influence mosquito breeding and disease transmission cycles.
    For instance, SST can serve as an early indicator of potential increases in regional temperatures and humidity levels, which are critical factors for mosquito population dynamics.
    ENSO phases, particularly El Niño, can be predictive of regional weather anomalies, such as increased rainfall in South America, which might lead to spikes in mosquito populations and extended disease transmission seasons.
    Incorporating PDO into the predictive models can add value by providing insight into broader climate trends that affect decadal weather patterns, influencing the longevity and intensity of mosquito breeding conditions over longer timescales.
    By integrating these climate indices into epidemiological models, public health officials can develop forecasts for the peak timing and duration of MBD outbreaks with greater confidence.
    Also, \citet{choi2022seasonal} shed light on the importance of analyze the relative contributions of internal model dynamics and external forcings to prediction skill, specially in wide time windows, where the importance dynamics significantly changes between these factors and \citet{foster2020bayesian} showed the importance in regularization by using constraints, such as fluctuation-dissipation theorem, to avoid overfitting. 
    
    The study by \citet{rader2023optimizing} introduced a approach to improve analog climate forecasting by utilizing a spatially-weighted mask derived through an interpretable neural network
    This method enhances the selection of analogs—historically similar climate states used to forecast future conditions—by applying a learned mask that emphasizes critical geographic regions, thus refining prediction accuracy.
    Mathematically, the approach involves multiplying the state of interest (SOI) and potential analogs by the weighted mask and then computing the mean squared error (MSE) between these weighted maps to assess similarity.
    The mask prioritizes areas crucial for accurate predictions while diminishing less relevant regions, effectively focusing the forecast model on the most predictive elements of the climate data.
    \citet{rader2023optimizing} employed the sea surface temperature (SST) from the Max Planck Institute for Meteorology Grand Ensemble (MPI-GE), which provides a comprehensive dataset with 100 ensemble members, each simulating 156 years of historical climate.
    The forecasting skill of the analogs selected using the learned mask was evaluated against traditional methods using a perfect model approach, where the same climate model data used for training the mask is employed to predict future states.
    This setup tested the efficacy of the weighted mask in a controlled environment, ensuring the improvements noted are due to the method itself rather than external variables. The results showed that the analogs selected with the weighted mask provide more skillful forecasts than those chosen through globally uniform weighting, underscoring the potential of this method to significantly advance seasonal-to-decadal climate forecasting.

    The study by \citet{rader2023optimizing} presents an innovative approach for enhancing the reliability of weather predictions, which could complement existing methodologies described by \citet{choi2022seasonal} and \citet{foster2020bayesian}.
    Also, its focus on optimizing seasonal-to-decadal analog forecasts using a learned spatially-weighted mask, refining the selection of analogs—historically similar social and climate states—by emphasizing crucial geographic regions.
    Moreover, \citet{rader2023optimizing} propose a skill data reduction methodology that proves beneficial for identifying analogous social and meteorological conditions across multiple regions.
    By analyzing the significance of various covariates on the magnitude of Mosquito Borne Disease incidence.
    It also suggests a strategy for enriching the training dataset with data from similar regions, thereby enhancing the predictive model's robustness and applicability. 

    The paper \citet{moulds2023skillful} presents an approach to forecasting abnormal streamflow events in the UK on a decadal scale using a statistical-dynamical framework.
    A large ensemble from the Coupled Model Intercomparison Project Phases 5 and 6 (CMIP5-6), providing monthly aggregate simulations of temperature and precipitation is used to model streamflows, particularly during the critical winter months (December to March).
    They employ generalized additive models for location, scale and shape (GAMLSS) to fit a Gamma distribution (log link) to predict the 95th percentile of daily winter streamflow (Q95), incorporating the predictors (temperature and precipitation).
    This model allows for adjustments based on temporal variability in climate data while maintaining the shape parameter as constant.
    The validation is carried out with a cross-validation procedure tailored for time-series data with continuous ranked probability score (CRPS) and the continuous ranked probability skill score (CRPSS) metrics, which assess the accuracy and skill of the forecasts relative to a simple climatological model.
    
    Just as the flood prediction model combines statistical and dynamical methods to forecast streamflow, a similar approach could be used to model mosquito-borne disease modeling.
    Knowing that MBD incidence can be influenced by climatic factors, this strategy can be employed to forecast abnormal incidence in endemic areas.
    Besides, the use of ensemble climate models to generate forecasts of temperature, precipitation and humidity can then be used to model mosquito population dynamics and disease transmission, accounting for the variability and uncertainty inherent in long-term forecasts.

    \citet{turchin20202010} revisits predictions made a decade ago about growing socio-political instability in the United States and Western Europe based on structural-demographic theory (SDT).
    Developed by pioneers such as Jack Goldstone and refined by Peter Turchin, SDT posits that societal stability is influenced by demographic and structural factors, evaluated through a mathematical model that quantifies mass mobilization potential (MMP), elite competition (measured as Elite Mobilization Potential, EMP) and state fiscal health (assessed as State Fiscal Distress, SFD).
    Each factor is derived from data: MMP factors in economic indices like inverse relative wage and youth demographics; EMP considers the ratio of elite incomes compared to GDP per capita and the density of elites relative to the population; SFD is gauged by metrics like government debt to GDP ratios and public trust in government.
    These components are integrated into a computational model, with the Political Stress Indicator (PSI or $\Psi$) computed as the product of MMP, EMP and SFD, to predict socio-political instability.
    In their assessment, the authors utilized comprehensive datasets from the Cross-National Time-Series Data Archive, covering events from 1815 to the present, which provided quantifiable measures of instability such as anti-government demonstrations and riots across several major Western countries.
    This extensive historical data helped to underpin and validate the model's projections for the 2010-2020 decade.
    Validation of the SDT-based forecasts was conducted by comparing these empirical socio-political instability indicators against the model’s predictions, demonstrating a significant correlation between the predicted increase in instability and the actual data observed during the 2010-2020 decade.
    This validation not only affirmed the accuracy of the SDT approach but also underscored its utility in understanding and forecasting complex societal dynamics over significant time periods.
   
    The concepts from SDT may give some insights on the influence of social, political, economic and public health variables to forecast the incidence and spread of diseases like malaria, dengue, Zika and West Nile virus over a decade.
    Key elements include MMP focusing on factors like economic distress (poverty rates, income inequality and employment rates), urbanization and demographic profiles; EMP examining political stability and elite competition as stability of governance structures impacts on public health policy and crisis management; and SFD, especially in relation to healthcare spending and public trust in government, that can hinder the effectiveness of public health campaigns and compliance with prevention measures.
    Incorporating these socio-political dimensions with historical and current epidemiological data on disease incidence and public health infrastructure will provide a foundational dataset.
    Computational models using differential equations can dynamically simulate these interactions, while statistical and machine learning models can identify predictive relationships.

\section{Conclusions and Open Research Questions}

    This review explored various long-term predictive modeling research for mosquito-borne diseases (MBDs), including those short-term predictive modeling and related disciplines or with greater expertise in long-term modeling, including meteorology and sociology, that could provide insights for MBDs comprehension. 
    It has been observed that the complexity of MBD transmission dynamics has a significant influence from climatic factors such as temperature and variations, as well as sociodemographic variations such as migration, urbanization patterns and population density, but not in linear terms. 
    Thus, its usage is not straightforward, and if not used with the appropriate caution, they could even diminish the predictive skill of the mathematical models and statistical methods as evidenced by various studies.
    
    Despite substantial progress, there are only a few studies focused on the development of decadal predictive models for MBDs, which is a reflex of the great difficulty in tackling the MBD forecast problem, even for short-term predictions. 
    Additionally, the majority of recent models focus predominantly on temperature-related variables, while the impacts of other climatic factors like humidity and wind speed and socio-demographic elements like population density, mobility and land use changes are less explored, being its role not completely understood. 
    Also, key questions like how we can effectively combine data from climate change projections and human demographic trends to make accurate predictions over longer periods.
    
    For future research, it is recommended to:
    \begin{itemize}
        \item Develop integrated models that combine data from climate change scenarios, human demographic changes and vector ecology to provide more accurate and region-specific forecasts for MBD incidence over longer time scales.
        \item Incorporate machine learning techniques to handle large datasets incorporating all available environmental evidence, to improve the robustness of predictive models, enabling them to adapt to new data and evolving disease dynamics.
        \item Analyze the patterns of how often abnormal outbreaks occur through the years and identify the most important factors that causes them.
        \item Eliminate predictors that, despite being important in short-term models, lose their influence in the long term and identify those factors that will only be important when their effects are accumulated over long periods.
    \end{itemize} 

\section*{Conflict of Interest}
The authors declare no conflicts of interest related to this work.
\section*{References}
\bibliographystyle{plainnat}
\bibliography{main}

\end{document}